\documentstyle[sprocl,epsfig,color]{article}

\def\PRD{{\em Phys. Rev.} D}

\def\be{\begin{equation}}
\def\ee{\end{equation}}
\def\bea{\begin{eqnarray}}
\def\eea{\end{eqnarray}}

\definecolor{Black}{named}{Black}
\definecolor{Blue}{named}{Blue}
\definecolor{Red}{named}{Red}
\definecolor{Green}{named}{ForestGreen}
\definecolor{Black}{named}{Black}
\definecolor{Olive}{named}{OliveGreen}
\definecolor{Royal}{named}{RoyalBlue}
\definecolor{Orange}{named}{YellowOrange}
\definecolor{Yellow}{named}{Goldenrod}
\definecolor{Cornblue}{named}{CornflowerBlue}
\definecolor{Lila}{named}{DarkOrchid}

\begin{document}

\hfill hep-ph/0410118

\hfill IPPP-04-51

\hfill DCPT-04-102

\title{MOTIVATION FOR POLARISED $E^-$ AND $E^+$ BEAMS\footnote{Summary of talks given at 
the Polarisation and Electroweak sessions at LCWS04.}}

\author{GUDRID MOORTGAT-PICK}

\address{IPPP, University of Durham, DH1 3LE Durham, UK}


\maketitle\abstracts{A future Linear Collider is well suited
for discovering physics beyond the Standard Model, for revealing the structure
of the underlying physics
 as well as for performing high precision tests of the Standard Model. 
The use of polarised beams will be one of
the powerful tools for reaching these goals.
This paper highlights some recent studies
of having both beams simultaneously polarised which 
allows to use longitudinally as 
well as transversely polarised beams. 
}

\section{Introduction And Overview}
The International Linear Collider (ILC) in the energy range up to about 1 TeV
has a large potential for the discovery of new particles  
and, due to its clear signatures, is perfectly suited for the 
precise analysis of
physics beyond the Standard Model (SM) as well as for testing the SM with an unprecedented
accuracy \cite{TDR}. An option for the ILC
is to operate in the GigaZ mode, i.e. to run with very high luminosity
at the $Z$ and $WW$ threshold. This will enable the most sensitive 
tests of the SM ever made.
Precision studies in this energy range, in
particular in the electroweak sector, allows to detect even marginal traces of 
deviations from the SM predictions. 
This provides sensitivity to the effects of heavy states of New Physics (NP) even
if they might not be directly produced at the LHC or the first energy phase of the ILC. 

An important tool of the ILC
is the use of polarised beams.
Already in the base line design it is foreseen to polarise the electron
beam. One expects a high polarisation degree between 80\% and 90\%, see e.g.\cite{yamamoto}.
Having both beams polarised (concerning possible designs for polarising $e^+$ at a LC,
see\cite{schweizer,ts-omori})  leads to
several additional advantages:
direct analysis of the interaction structure of NP,
 increased sensitivity to non-standard as well as SM couplings,
improved accuracy in measuring the polarisation (see also\cite{moenig2}), 
higher effective polarisation $P_{eff}=(P_{e^-}-P_{e^+})/(1-P_{e^-}P_{e^+})$,  
and enhanced signal and suppressed background rates for specific
processes with a suitably chosen polarisation configuration; in addition
also the possibility is provided to use transversely polarised beams
 (for further overview reports see e.g. \cite{Steiner,Omori}). 

A comprehensive report is in preparation in which 
the physics arguments for having both beams polarised are summarised
in detail. 
In the report also an 
overview about possible technical designs for 
producing polarised beams and for measuring the polarisation\cite{Power} is given.
In the following we briefly summarise  
some recent studies with  
longitudinally as well as transversely polarised beams.

\begin{figure}
\begin{center}
     \vspace*{.2cm}
\begin{picture}(7,7)
\setlength{\unitlength}{1cm}
\put(-6,0){\mbox{\includegraphics[height=.25\textheight]{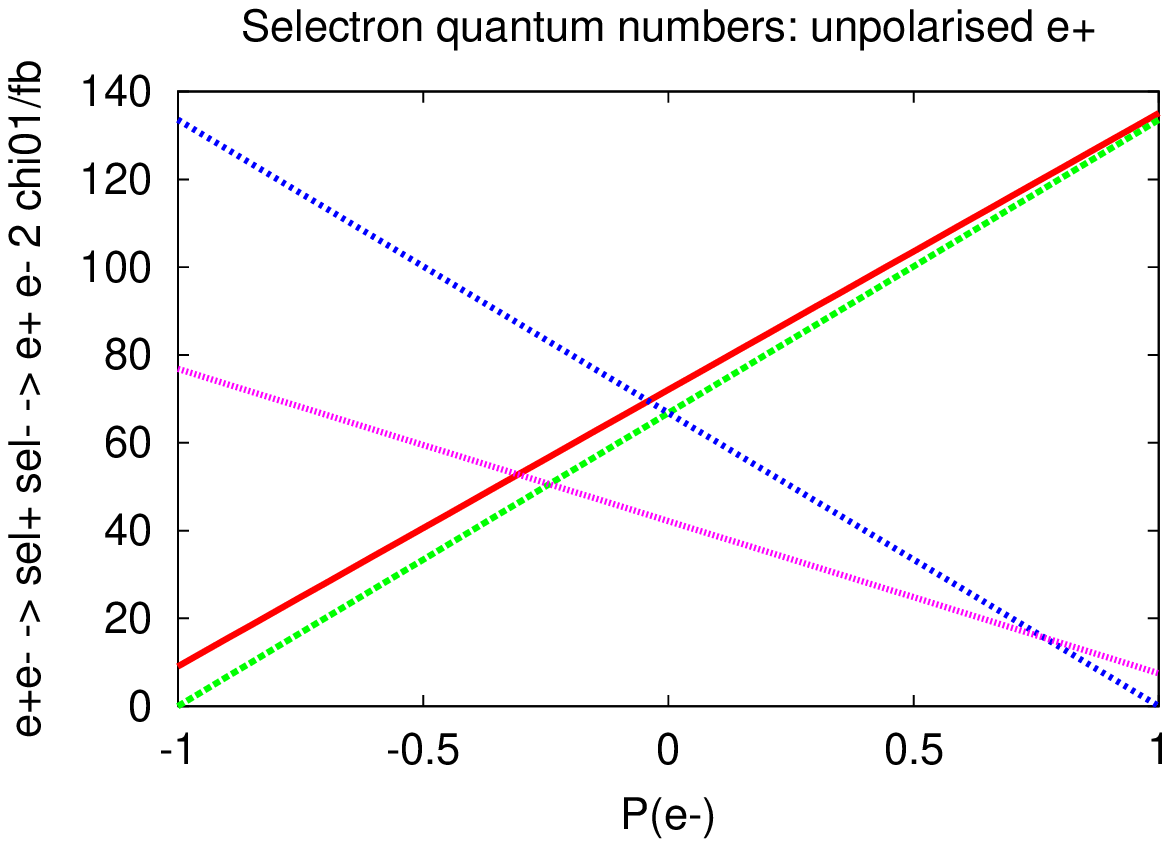}}}
\put(-.9,2.3){\small \makebox(0,0)[br]{{\color{Green}
$\tilde{e}^+_{L} \tilde{e}^-_{R}$}}}
\put(-4.1,1.7){\small \makebox(0,0)[br]{{\color{Lila}
$\tilde{e}^+_{L} \tilde{e}^-_{L}$}}}
\put(-4.1,2.6){\small \makebox(0,0)[br]{{\color{Blue}
$\tilde{e}^+_{R} \tilde{e}^-_{L}$}}}
\put(-1.9,2.8){\small \makebox(0,0)[br]{{\color{Red}$
\tilde{e}^+_{R} \tilde{e}^-_{R}$}}}
\put(-6,4.2){\tiny $\sqrt{s}=500$~GeV}
\put(.5,0){\mbox{\includegraphics[height=.25\textheight]{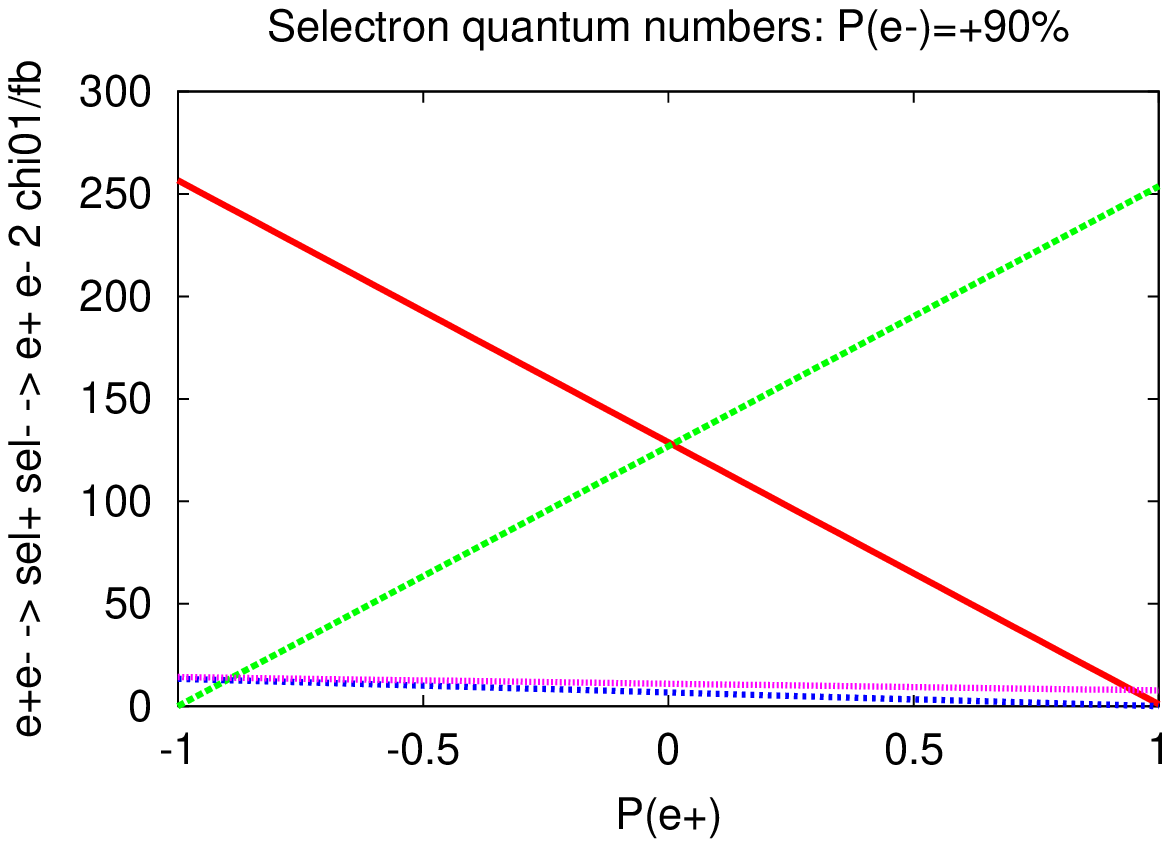}}}
\put(5.7,2.2){\small \makebox(0,0)[br]{{\color{Green}
$\tilde{e}^+_{L} \tilde{e}^-_{R}$}}}
\put(2.5,2.3){\small \makebox(0,0)[br]{{\color{Red}$
\tilde{e}^+_{R} \tilde{e}^-_{R}$}}}
\put(.5,4.2){\tiny $\sqrt{s}=500$~GeV}
\end{picture}\vspace{1cm}
\end{center}\vspace{-1.5cm}
\caption{\label{fig-sel-var-el} Separation of the selectron pair
$\tilde{e}_L^-\tilde{e}^+_R$ in $e^+ e^-\to \tilde{e}^+_{L,R}
\tilde{e}^{-}_{L,R}$ may not be possible with 
electron polarisation only (left); if, however, both beams are polarised the
$RR$ configuration separates the pairs and the association of the selectrons to the chiral 
quantum numbers can be exerimentally tested$^{8,9}$.}
\end{figure}
\section{Both Beams Longitudinally Polarised}
Concerning the possible helicity combinations in the interactions one has to distinguish two cases:
\begin{itemize}
\item[a)] in annihilation diagrams the helicities of the incoming beams
are coupled to each other, whereas
\item[b)] in scattering diagrams the helicities of both incoming beams are
directly coupled to the final particles.
\end{itemize}
In case a) in the SM only the recombination into a vector
particle with the total angular momentum $J=1$ is possible (in the limit 
$m_e\to 0$), i.e.\
both beams have to carry opposite sign of helicities. 
Only models of NP (in this limit) might allow to produce also scalar 
particles, so that
$J=0$ would be allowed, which results in same sign helicities
of the incoming beams. In case b) the scattering diagrams could result in a vector,
fermionic or scalar particle; the helicity of the incoming
particle is directly coupled to the vertex and is independent of
the helicity of the second incoming particle. Therefore all
helicity configurations are possible.\\[-.3em]

\noindent {\bf a) Verifying quantum numbers of new particles}\\[.3em]
We show how crucial it may be to have both beams polarised and choose
one representative example in Supersymmetry (Susy), which is one of the best motivated 
theory candidates for physics beyond the SM.
At the LHC and the ILC, one has -- after observing signals of 
new physics -- to prove for instance, that it is indeed Susy, i.e.\  to verify
the Susy predictions: in particular, Susy particles have to carry the
same quantum numbers as their SM
partners (with the exception of the spin which differs by half a unit).
Thus Susy transformations associate chiral (anti)fermions to
scalars, i.e. $e^-_{L,R}\leftrightarrow \tilde{e}^-_{L,R}$ but
$e^+_{L,R}\leftrightarrow \tilde{e}^+_{R,L}$. In order to prove
this association a simultaneous polarisation of both beams is decisive
\cite{Power,Bloechi}.  The process $e^+e^-\to \tilde{e}^+\tilde{e}^-$
occurs via $\gamma$ and $Z$ exchange in the annihilation
channel and via
neutralino $\tilde{\chi}^0_i$ exchange in the scattering channel. The
association can be directly tested only in the  latter channel. The
use of polarised beams serves to separate both channels. 

We demonstrate this in our example, Fig.~\ref{fig-sel-var-el}, 
by isolation of the pair $\tilde{e}^+_L
\tilde{e}^-_R$ applying right-handed polarisation of both beams 
($RR$ configuration, i.e. $P_{e^-}>0$ and $P_{e^+}>0$). The selectron masses are close together,
$m_{\tilde{e}_L}=200$~GeV, $m_{\tilde{e}_R}=195$~GeV, 
 so that both $\tilde{e}_{L}$, $\tilde{e}_R$ show the same decay,
$\tilde{e}_{L,R}\to \tilde{\chi}^0_1 e$.
Even extremely high electron polarisation, $P_{e^-}\ge +90\%$, is not  
sufficient to
disentangle the pairs $\tilde{e}_L^+\tilde{e}_R^-$ and
$\tilde{e}_R^+ \tilde{e}_R^-$ and to test their association to the chiral quantum numbers, 
since both cross sections are very close to each other, Fig.~\ref{fig-sel-var-el} (left). 
Only with right-handed polarisation of both beams 
the pair $\tilde{e}^+_L \tilde{e}^-_R$ is clearly separated, see Fig.~\ref{fig-sel-var-el} (right).
In addition, all SM background events, e.g.\ from $W^+W^-$, are
strongly suppressed with this $RR$ configuration.

\begin{figure}[t]
\begin{picture}(8,5)
\setlength{\unitlength}{1cm}
\put(0,0){\epsfig{file=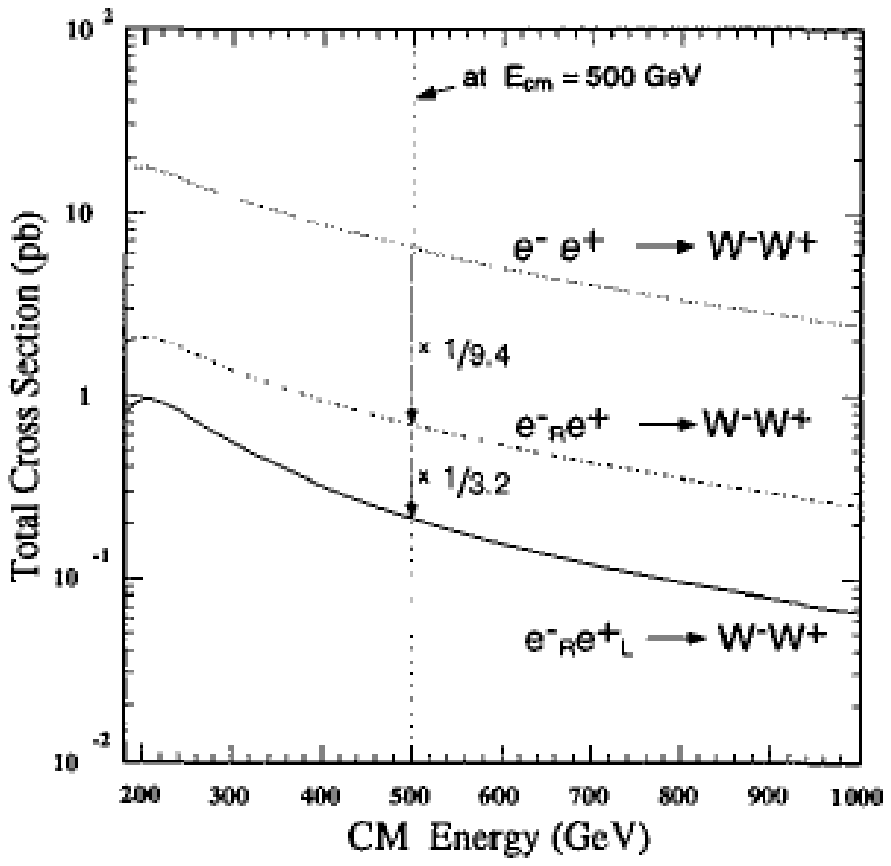,scale=.5,width=.33\textheight,height=.25\textheight}}
\put(6.2,.1){\mbox{\includegraphics[height=.24\textheight,width=.33\textheight]{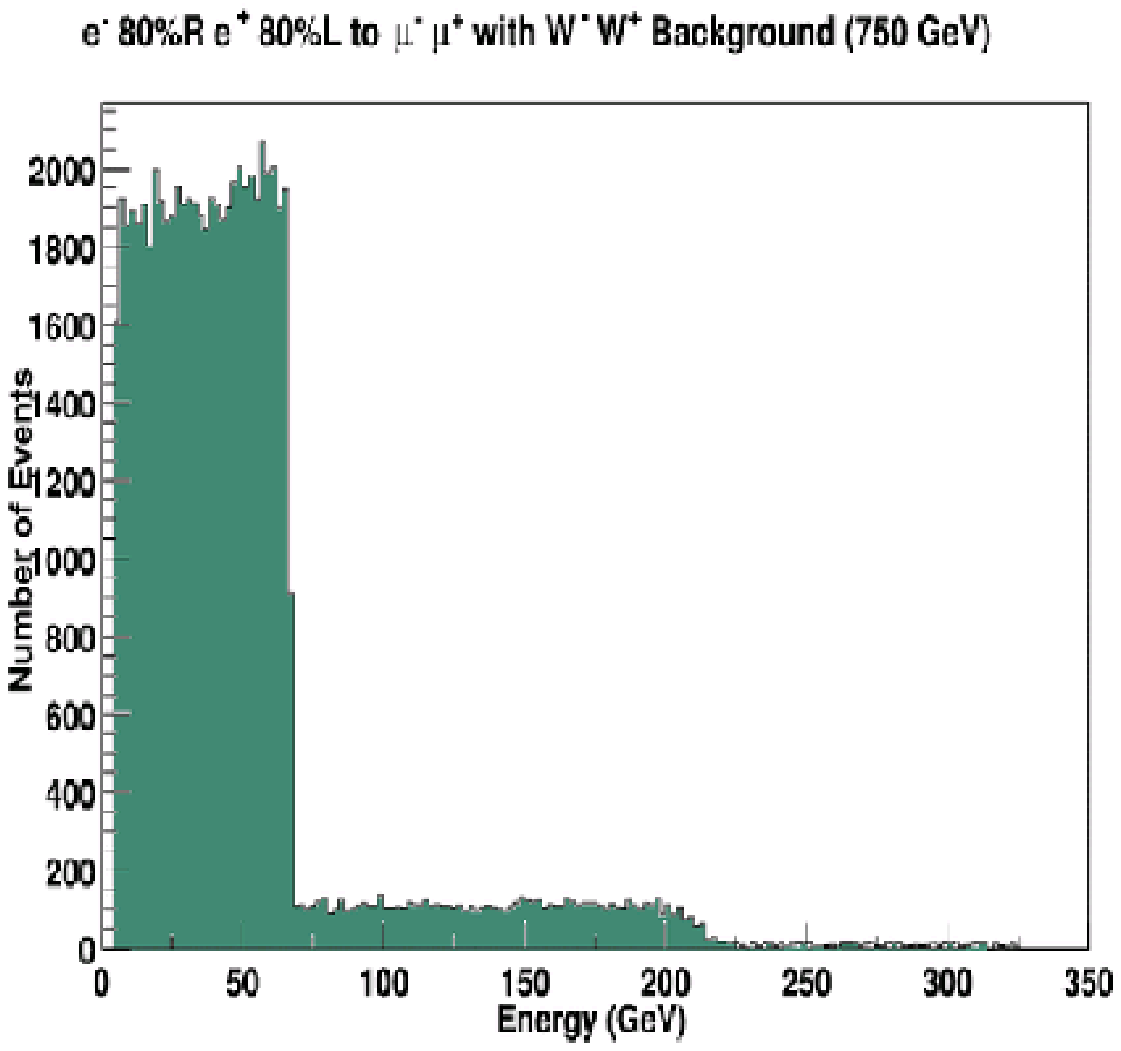}}
}
\end{picture}
\vspace{-.3cm}
\caption{Left: Cross sections for $e^+e^-\to W^+W^-$ for unpolarised beams and for
$P_{e^-}=+90\%$, $P_{e^+}=0\%$ and $P_{e^-}=+90\%$, $P_{e^+}=-80\%$ as function of $\sqrt{s}^{7}$.
Right: Energy spectrum of muons from $\tilde{\mu}_{L,R}$ decays into $\mu\nu$ final states including the
$W^+W^-$ background for the beam polarisations 
$P_{e^-}=+80\%$, $P_{e^+}=-80\%$ for
$\sqrt{s}=750$~GeV$^{10}$.
\label{sigmaWW} }
\end{figure}

\vspace*{.3cm}
\noindent {\bf b) Gain in statistics and suppression of backgrounds}\\[.3em]
In many cases of NP models, e.g.\ in Susy models, the predicted
production cross sections of the new particles could be very small and
the use of a suitable polarisation of both beams may be decisive for observing the signal
(and saving running time).  Simulations have been done to
suppress SM background processes with the help of polarised beams in
new physics searches, where $W^+W^-$ production presents one of the
worst SM backgrounds. However, it can be easily suppressed with
right-handed electron/left-handed positron beams. The reduction of the $W^+W^-$ cross section is
$\sigma^{pol}/\sigma^{unpol}=$0.2 (0.1) for $P_{e^-}=+80\%$,
$P_{e^+}=0$ ($P_{e^-}=+80\%$ and $P_{e^+}=-60\%$); see
Figure~\ref{sigmaWW} (left)\cite{Omori} for $P_{e^-}=+90\%$,
$P_{e^+}=-80\%$.

The background suppression may be decisive for the detection of new
particles and the precise determination of their properties, e.g. the
accurate measurement of $m_{\tilde{\mu}}$ in the continuum.  In the
case of $\tilde{\mu}^+\tilde{\mu}^-$ production we have only
annihilation via $\gamma$ and $Z^0$ exchange, therefore the initial
beam configurations $LR$ and $RL$ are favoured.  The predominant
background for the signal is $W^+W^-$ production.  In \cite{uriel} an
example is given, in which the masses are $\tilde{\mu}_R=178.3
\mbox{ GeV},\quad \tilde{\mu}_L=287.1 \mbox{ GeV}$.  In
Fig.~\ref{sigmaWW} (right) the expected muon energy distribution for
an integrated luminosity of 500~fb$^{-1}$ at $\sqrt{s}=750$~GeV and
the polarisation configuration $P_{e^-}=+80\%$, $P_{e^+}=-80\%$ is
shown. The background from $W^+W^-$ decaying into the $\mu \nu$ final
state is included. Only with polarised $e^-$ and $e^+$ beams both
edges, at around 65~GeV and at around 220~GeV, can be clearly
reconstructed. The slepton masses can be determined in the continuum
up to a few GeV, which is important to test e.g.\ slepton
non-universality \cite{Baer}.  The mass
reconstruction is even more involved in the case of selectrons. The $e^+e^-$ energy
spectra subtraction technique has been successfully applied to remove
SM background \cite{uriel}.  Polarisation of both beams is needed to
guarantee sufficient statistics.  The mass measurements in the
continuum are very important to outline strategies for further
threshold scans for specific particles, cf.\cite{TDR}.

In \cite{Hesselbach1} it has been shown that the use of polarised beams could in addition be needed to 
distinguish between different Susy models, the MSSM and the NMSSM. 
Susy models often lead to a large number of additional free parameters and  
new sources for CP violation.
Small effects are expected in most
CP violating Susy processes. Suitable observables for uniquely resolving the CP structure of the 
underlying physics
are T-odd asymmetries. Using polarisation of both beams 
may be important to enhance the cross sections 
and to get sufficient statistics \cite{Power,Hesselbach2}. 

A further example for the importance of having both beams polarised in order to suppress SM background
for new physics searches
can be seen in the process $e^+e^-\to \gamma G$, the direct search for gravitons. The predominant 
background is
$e^+e^-\to \gamma \nu_e \bar{\nu}_e$. Given the near maximal polarisation asymmetry of the background,
polarised beams are extremely effective in suppressing the background:
the ratio $S/\sqrt{B}$ increases by a factor 2.1
when using $(P_{e^-},P_{e^+})=(+0.8,0)$ and by a factor 4.4 when using 
$(+0.8,-0.6)$, see\cite{wilson}. 
The reach of the LC in the quest for evidence of extra dimensions is thus extended
and leads to a similar discovery reach 
as at the LHC in a, however, more model-independent way\cite{wilson}.\\[-.3em]

\begin{figure}[t]
\begin{minipage}{5cm}
\begin{picture}(10,4)
\setlength{\unitlength}{1cm}
\put(0,1){\epsfig{file=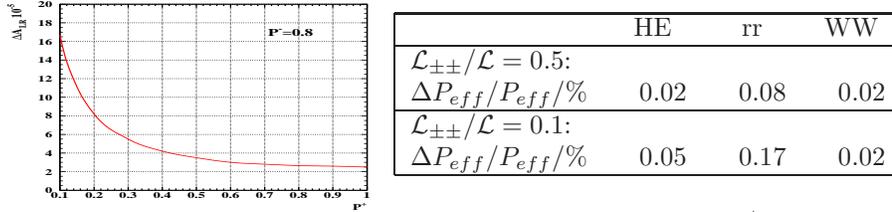,scale=.25,width=.33\textheight,height=.2\textheight}}
\end{picture}
\end{minipage}\hspace{.1cm}
\begin{minipage}{5cm}
{\vspace*{-5cm}
\begin{tabular}[c]{|cccc|}
      \hline
      & HE         & rr &WW\\ \hline
       \multicolumn{4}{|l|}{${\cal L}_{\pm\pm} /{\cal L} = 0.5$:} \\
      $\Delta P_{eff}/P_{eff}$/\%  & $\phantom{-}0.02$ & $\phantom{-}0.08$ & 
      $\phantom{-}0.02$\\
    \hline 
          \multicolumn{4}{|l|}{${\cal L}_{\pm\pm} /{\cal L} = 0.1$:} \\  
           $\Delta P_{eff}/P_{eff}$/\%  & $\phantom{-}0.05$ & $\phantom{-}0.17$ & 
      $\phantom{-}0.02$\\  
      \hline
    \end{tabular}
    }
\end{minipage}\vspace{-1.5cm}
\caption{Left: The statistical error on the left--right asymmetry
$A_{LR}$ of $e^+ e^-\to Z\to \ell \bar{\ell}$ at GigaZ
as a function of the positron
polarisation $P(e^+)$
for fixed electron polarisation $P_{e^-}=\pm 80\%$$^5$; Right: Relative polarisation error of $P_{eff}$
using the 
Blondel scheme for 
      $\sqrt{s}=340$~GeV, ${\cal L}=500$~fb$^{-1}$, $P_{e^-}=0.8$, $P_{e^+}=0.6$, cf.$^{5}$
      (HE = High energy events from $e^+e^-\to f \bar{f}$, 
      rr = radiative return via $e^+e^-\to Z\gamma\to f \bar{f}\gamma$, WW =  W-pair production,
      ${\cal L}_{\pm\pm} /{\cal L} = 0.5$(0.1) of luminosity spent on same sign
      beam polarisation). \label{alr-blondel}}
\end{figure}

\noindent {\bf c) High precision SM tests at GigaZ}\\[.3em]
In the SM the left--right asymmetry $A_{LR}$ of $e^+e^-\to Z^0\to f \bar{f}$ depends only on the effective
leptonic weak mixing angle:\\[-1em]
{\small
\begin{equation}
A_{LR}=\frac{2(1-4 \sin^2\Theta^l_{eff})}{1+(1-4 \sin^2\Theta^l_{eff})^2}.
\label{eq_ew1}
\end{equation}
}
The statistical power of the data sample can be fully exploited only when
$\delta(A_{LR}(pol))<\delta(A_{LR}(stat))$.
For $10^8-10^9$ Z's this occurs when
$\delta(P_{eff})\le 0.1\%$. In this limit
$\delta(\sin^2\theta_{eff})\sim 10^{-5}$, which is more than an order of magnitude
smaller than the present value of this error. Thus it will be crucial to
minimise the error in the determination of the polarisation.
The desired precision  is attainable with
the Blondel Scheme: it is not necessary to
know the beam  polarisation itself with such an extreme accuracy, since
$A_{LR}$ can be directly expressed via polarised cross sections\cite{moenig2}.
Fig.~\ref{alr-blondel} (left)
shows the statistical error on $A_{LR}$ as a function of the positron
polarisation for $P_{e^-}=80\%$\cite{moenig2}.
The Blondel scheme also requires some
luminosity for the less favoured combinations (LL, RR). However, only about
10\% of running time will be needed for these combinations
to reach the desired accuracy for these high precision measurements, 
cf. Fig.~\ref{alr-blondel} (right)~\cite{moenig2}.

\begin{figure}[t]
\begin{minipage}{5cm}
\begin{picture}(4,6)
\setlength{\unitlength}{1cm}
\put(-.5,0){\mbox{
\includegraphics[width=3.2cm,angle=90]{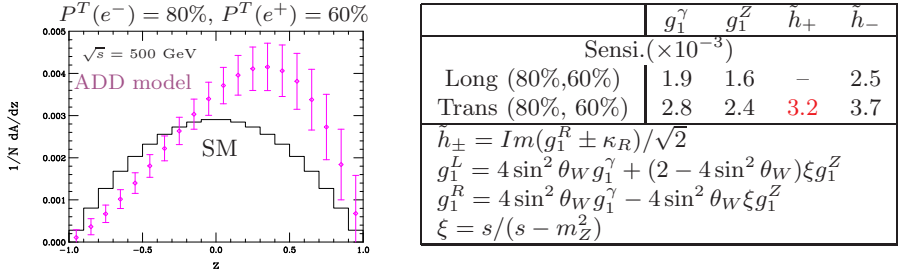}}}
\put(.6,2.8){\tiny $\sqrt{s}=500$~GeV}
\put(.3,3.3){\footnotesize $P^T(e^-)=80\%$, $P^T(e^+)=60\%$}
\put(.5,2.4){\footnotesize\color{Lila} ADD model}
\put(2.2,1.5){\small SM}
\end{picture}
\end{minipage}
\begin{minipage}{5cm}
{\small\vspace{-3.6cm}
\begin{tabular}{|c|cccc|}
\hline
 & $g_1^{\gamma}$ & $g_1^Z$ &
 $\tilde{h}_{+}$ & $\tilde{h}_{-}$ \\ \hline
\multicolumn{5}{|c|}{Sensi.($\times 10^{-3}$)}\\
Long (80\%,60\%) & 1.9 & 1.6 & -- & 2.5\\
Trans (80\%, 60\%) & 2.8 & 2.4  & {\color{Red}3.2} & 3.7\\
\hline
\multicolumn{5}{|l|}{$\tilde{h}_{\pm}=Im(g_1^R\pm\kappa_R)/\sqrt{2}$}\\
\multicolumn{5}{|l|}{$g_1^{L}=4 \sin^2\theta_W g_1^{\gamma}+(2-4\sin^2\theta_W)\xi g_1^Z$}\\
\multicolumn{5}{|l|}{$g_1^{R}=4 \sin^2\theta_W g_1^{\gamma}-4\sin^2\theta_W \xi g_1^Z$}\\
\multicolumn{5}{|l|}{$\xi=s/(s-m_Z^2)$}\\
\hline
\end{tabular}
}
\end{minipage}\vspace{-.3cm}
\caption{Left: Differential azimuthal asymmetry 
distribution for $e^+e^-\to c \bar{c}$ 
at a 500 GeV LC assuming a luminosity of $500~fb^{-1}$. The histograms are 
the SM predictions while the data points assume the ADD model with $M_H=1.5$ 
TeV$^{15}$; Right: 1$\sigma$ statistical errors on the 
real and the imaginary parts of TGCs ($CP$ conserving)
in the presence of general anomalous
couplings at $\sqrt{s}=$ 500 GeV with different beam 
polarisations$^{16}$.\label{transverse}}
\end{figure}

\section{Both Beams Transversely Polarised:\\ 
Searches For Gravitons, TGCs And CP Violation}
One elegant tool for new physics searches that becomes only available if both 
$e^-$ and $e^+$ beams are polarised at the LC are
transversely polarised beams. In that case the cross sections can generally be written as:\\[-.8em]
\[
\sigma=(1-P_{e^+}P_{e^-})\sigma_{unp}\hspace*{.1cm}+\hspace*{.1cm}
(P_{e^-}^L-P_{e^+}^L)\sigma_{pol}^L
\hspace*{.1cm}+\hspace*{.1cm}P_{e^-}^T P_{e^+}^T \sigma_{pol}^T.\]

It has been worked out in~\cite{Rizzo} that with transversely polarised beams
a unique distinction between effects of graviton exchange and
of 'conventional' contact interactions is possible. The interference between 
SM and spin-2 exchange amplitudes 
shows a general difference in the $z$-dependence ($z=\cos\theta$) of the terms which are  
sensitive to the azimuthal angle. In particular, the existence of the odd $z$ 
contributions is clearly a signal for spin-2 exchange, see Fig.~\ref{transverse} (left).  

It has been shown in \cite{Nagel} that transversely polarised beams may also be crucial for
the determination of a specific triple gauge coupling, $\tilde{h}_{+}$, that is not accessible 
with only longitudinally polarised beams, see Fig.~\ref{transverse} (right).

Furthermore it
has been shown in \cite{Anant} that transversely polarised beams also enhance the sensitivity
to CP-odd observables at leading order from interferences of (pseudo-) scalar or tensor currents
with $\gamma$ and $Z$  channels in general inclusive processes, $e^+e^-\to A+X$.


\section{Summary}
The ILC in the TeV range with its 
clean initial state of $e^+ e^-$ collisions is 
ideally suited to search for New Physics. It allows to precisely determine quantities of 
the Standard Model as well as of New Physics and to reveal 
the structure of the underlying physics.
The use of polarised beams plays a decisive role in this context.
The simultaneous polarisation of both beams, longitudinally as well as transversely, 
significantly expands the physics potential, e.g.\ for verifying and determining  
the properties of new particles, 
for increasing signal rates and suppressing background processes, and 
for providing higher sensitivity to
non--standard couplings and to CP effects of new physics.

\vspace*{-.3cm}

\end{document}